\documentclass[journal]{IEEEtran}  

\IEEEoverridecommandlockouts                              

\usepackage{amsmath} 
\usepackage{amssymb}  
\usepackage{amsthm}
\usepackage{mathtools}
\usepackage{algorithm}
\usepackage{algpseudocode}
\usepackage{gensymb}
\usepackage{hyperref}
\usepackage{siunitx}
\usepackage{pifont}
\usepackage{cite}
\usepackage{listings}  
\usepackage{graphicx}
\usepackage{tabularx}
\usepackage{multirow}
\usepackage{tikz}
\usepackage{pgfplots}
\graphicspath{ {./graphics/} }
\usepackage[caption=false]{subfig}
\usepackage{booktabs}
\usepackage{xcolor}

\DeclareMathOperator*{\argmin}{argmin}

\definecolor{codegreen}{rgb}{0,0.6,0}
\definecolor{codegray}{rgb}{0.5,0.5,0.5}
\definecolor{codepurple}{rgb}{0.58,0,0.82}
\definecolor{backcolour}{rgb}{0.96,0.97,0.97}

\lstdefinestyle{mystyle}{
    backgroundcolor=\color{backcolour},   
    commentstyle=\color{codegreen},
    keywordstyle=\color{magenta},
    numberstyle=\tiny\color{codegray},
    stringstyle=\color{codepurple},
    basicstyle=\ttfamily\footnotesize,
    breakatwhitespace=false,         
    breaklines=true,                 
    captionpos=b,                    
    keepspaces=true,                 
    numbersep=5pt,                  
    showspaces=false,                
    showstringspaces=false,
    showtabs=false,                  
    tabsize=2
}

\pgfplotsset{compat=1.18}

\title{CommonPower: A Framework for Safe \\ Data-Driven Smart Grid Control}

\author{Michael Eichelbeck, Hannah Markgraf, and Matthias Althoff%
\thanks{This work was partially supported by the German Research Foundation (grant no. 458030766) and the Bavarian Research Foundation project STROM (Energy - Sector coupling and microgrids, AZ-1473-20). The authors are with the School of Computation, Information and Technology, Technical University of Munich, Germany. \newline
E-mail: {\tt\small michael.eichelbeck@tum.de}}%
}

\begin{document}

\maketitle
\thispagestyle{empty}
\pagestyle{empty}

\begin{abstract}

The growing complexity of power system management has led to an increased interest in reinforcement learning (RL). To validate their effectiveness, RL algorithms have to be evaluated across multiple case studies. Case study design is an arduous task requiring the consideration of many aspects, among them the influence of available forecasts and the level of decentralization in the control structure. 
Furthermore, vanilla RL controllers cannot themselves ensure the satisfaction of system constraints, which makes devising a safeguarding mechanism a necessary task for every case study before deploying the system. 
To address these shortcomings, we introduce the Python tool \textit{CommonPower}, the first general framework for the modeling and simulation of power system management tailored towards machine learning. Its modular architecture enables users to focus on specific elements without having to implement a simulation environment. 
Another unique contribution of CommonPower is the automatic synthesis of model predictive controllers and safeguards. 
Beyond offering a unified interface for single-agent RL, multi-agent RL, and optimal control, CommonPower includes a training pipeline for machine-learning-based forecasters as well as a flexible mechanism for incorporating feedback of safeguards into the learning updates of RL controllers. 

\end{abstract}

\begin{IEEEkeywords}
Safe reinforcement learning, energy management, model predictive control, multi-agent systems, and forecast uncertainties.
\end{IEEEkeywords}

\section{Introduction}

\IEEEPARstart{I}{ncreasing} adoption of intermittent renewable energy generation and complex power demand patterns, e.g., from electrifying heating and mobility, challenge power system control. Reinforcement learning (RL) has emerged as a promising method as it does not require explicit model knowledge and can automatically adapt to changing parameters. RL controllers have successfully been demonstrated for Volt-Var control, frequency control, economic dispatch, and smart home energy management \cite{chen2022reinforcement}. Additionally, distributed control based on multi-agent reinforcement learning (MARL) provides a versatile data-driven approach largely using local information \cite{vazquez2019citylearn, pigott2022gridlearn, biagioni2021powergridworld}. 

RL controllers have demonstrated competitive performance compared to model predictive controllers (MPC) at significantly lower computational costs \cite{dasilva2022battery}. Furthermore, there is evidence that RL controllers can learn the implicit patterns of disturbances and thus outperform a nominal MPC in a setting with inaccurate forecasts \cite{zhang2020restoring}.
For all its promise, integrating RL controllers into smart grids faces a major challenge since vanilla RL cannot guarantee the satisfaction of system constraints. 

Safe RL is an active area of research with a large amount of existing literature \cite{garcia2015comprehensive, gu2022review, könighofer2022correct, bui2025critical}. While there exist many algorithms that satisfy constraints with high probability, the critical nature of power systems renders guaranteed constraint satisfaction a strict requirement for the real-world deployment of RL controllers. 
Such mechanisms are generally hand-crafted for individual case studies \cite{cui2022decentralized, vu2021safe, vazquez2020marlisa, bahrami2020deep}, which is a tedious task for practitioners and can be prohibitively hard for non-experts. Further, it does not allow for a general approach for passing feedback from the safety mechanism to the RL agent, which has been shown to influence control performance \cite{krasowski2022provably, markgraf2023safe}. 
A more generic approach is to establish safety guarantees based on simplified system models that enclose all possible behaviors considering parametric uncertainties and disturbances \cite{krasowski2022provably}.

Beyond the study of safeguarding mechanisms, there are two further aspects of smart grid control that are becoming increasingly relevant and motivate further research. 
Firstly, the adoption of local energy communities, virtual power plants, and supply/demand aggregators results in more and more distributed control settings combining different types of controllers \cite{uddin2023micogrids, alismail2021dc}.
Secondly, data-driven models are becoming the standard forecasting approach as they show significant promise to improve forecast accuracy and thus control performance \cite{amasyali2018review, foley2012current, wang2019review, das2018forecasting}.
Before presenting how our tool addresses these challenges in Section \ref{subsec:contributions}, we provide an overview of existing Python libraries for modeling power systems and interfacing RL agents.

\subsection{Related Work}\label{sec:related_work}

\textit{Andes\_gym} \cite{cui2022andes_gym} provides one single-agent RL environment for frequency and voltage control. It is based on the \textit{ANDES} library \cite{cui2020hybrid}, which features a symbolic modeling framework and optimized numerical simulations. \textit{Gym-ANM} \cite{henry2021open-source, henry2021reinforcement} targets economic dispatch use cases and notably includes an MPC as a baseline, which is, however, limited to the pre-defined device models. \textit{Grid2op} \cite{kelly2020reinforcement, marot2021learning} presents a framework for power grid management in which agents can control both the grid topology and power dispatch. Furthermore, it can model opponents that attempt to destabilize the system. \textit{PowerGym} \cite{fan2021powergym} is a library designed for Volt-Var control in distribution networks and utilizes the Python version of OpenDSS for solving provided system constraints. The tool \textit{python-microgrid} \cite{henri2020pymgrid} is a lightweight framework for economic dispatch in microgrids. \textit{RLGC} \cite{huang2019adaptive} is a library tailored for emergency control, e.g., generator dynamic braking or under-voltage load shedding using the Java tool InterPSS for simulation. Lastly, \textit{SustainGym} \cite{yeh2023sustaingym} is a collection of benchmarks for single-agent and multi-agent control covering five distinct energy management use cases.  

Toolboxes for MARL exist mainly for home energy management and economic dispatch applications. \textit{CityLearn} \cite{vazquez2019citylearn, nweye2025citylearn} provides both benchmark environments and baseline implementations for demand response of buildings using single-agent and multi-agent RL. This framework is extended by a power grid model in \textit{GridLearn} \cite{pigott2022gridlearn} so that tasks, such as voltage regulation, can be addressed. Both libraries focus on decentralized control of active storage components, such as thermal energy storage or batteries, and rely on pre-simulated heating and cooling demands. \textit{PowerGridWorld} \cite{biagioni2021powergridworld} provides a modular framework for modeling multi-agent scenarios, relying on OpenDSS for solving power flow.

All existing tools share the key limitation that they do not maintain a symbolic system model that can readily be exposed to controllers or safeguards. This disconnect requires users to implement model-based safeguarding or model-based control specifically for each case study, which becomes a very laborious task for large or heterogeneous systems. Device-level safety mechanisms are hard-coded and cannot readily provide feedback to controllers.
All tools either rely on a set of built-in, specifically structured components or external simulation tools that require complex tabular configuration files, limiting flexibility and ease of use.
Further, existing tools cannot simulate different types of controllers, such as RL-based, rule-based, or model-based, in the same system. This limits the possibility of modeling complex distributed control structures.
Lastly, of all investigated tools, only python-microgrid provides a generic forecaster interface, and none includes functionality for developing machine-learning-based forecasters. This makes it cumbersome to study the influence of forecaster accuracy and requires external tool support for the training of machine-learning-based forecasters.

\subsection{Contributions}\label{subsec:contributions}

Our Python library CommonPower\footnote{\url{https://github.com/TUMcps/commonpower}} closes the aforementioned gaps and addresses the need for a versatile tool facilitating the exploration of safe controllers in a large variety of use cases. As such, it provides a \textit{common} ground for researchers and practitioners in the area of data-driven smart grid control.
CommonPower contains the following main features:

\begin{itemize}
    \item \textbf{Modular architecture}: CommonPower has a highly modular approach in which power system entities, controllers, safeguards, forecasters, and data sources are abstracted as objects with clearly defined interfaces.
    \item \textbf{Flexible component modeling}: As a framework, CommonPower facilitates the modeling of arbitrary scenarios. The built-in models of buses, devices, or power flow can easily be extended, or entirely custom components can be implemented, and their symbolic model is automatically considered. CommonPower integrates several modeling utilities, e.g., for piece-wise continuous models, and supports the integration of external tools to simulate complex dynamics of individual devices. 
    \item \textbf{Adaptive RL safeguarding}: Since CommonPower maintains a symbolic representation of the system under study, model-based safeguarding approaches for RL can be derived automatically. The built-in implementation is based on a robust optimal control formulation of the system constraints and considers model uncertainties as well as disturbances. 
    \item \textbf{Built-in robust MPC}: CommonPower utilizes its symbolic representation to automatically synthesize a robust model predictive controller that can serve, e.g., as a credible baseline or as a basis for imitation learning.  
    \item \textbf{Unified RL interface}: CommonPower implements a unified gymnasium environment \cite{towers2023gymnasium} for single-agent and multi-agent RL. This facilitates the comparison of both paradigms and makes it possible to directly integrate any algorithm or library supporting the gymnasium interface. 
    \item \textbf{Heterogeneous distributed control structures}: Due to the modular design of CommonPower, different types of controllers can be combined in the same multi-agent system. Based on a highly flexible problem formulation, any controller can be mapped to an arbitrary number of controllable entities. 
    \item \textbf{Data-driven forecasting}: Beyond providing a generic forecaster interface, CommonPower implements a framework for training, evaluating, and tuning machine-learning-based prediction models.
    \item \textbf{Integration and documentation}:
    To facilitate integration into ongoing projects, a power grid import interface to the well-known library \textit{pandapower} \cite{thurner2018pandapower} is provided. Furthermore, CommonPower implements an interface to the library \textit{PyTupli} \cite{markgraf2025pytupli}, a toolbox for managing experience tuples for offline and continual learning.
    The code base is well documented and includes several tutorials to support user on-boarding\footnote{\url{https://commonpower.readthedocs.io/en/latest/}}. 
\end{itemize}

\subsection{Organization}

After introducing some notation and background (Sec. \ref{sec:preliminaries}), we formulate our high-level control problem and establish its corresponding RL environment (Sec. \ref{sec:problem_statement}). Afterward, we describe how power systems and forecasters are modeled in CommonPower (Sec. \ref{sec:modeling}), followed by presenting our unified approaches to control and safeguarding (Sec. \ref{sec:control}). Finally, we investigate several case studies (Sec. \ref{sec:experiments}) and conclude (Sec. \ref{sec:conclusion}). 

\section{Preliminaries}\label{sec:preliminaries}

\begin{figure*}[t]
\centering
    \subfloat[Two-stage control with balancing assets operated by an MPC and two different coalitions controlled by multi-agent RL.]{
        \label{fig:opt1}
        \centering
        \includegraphics[width=0.45\textwidth]{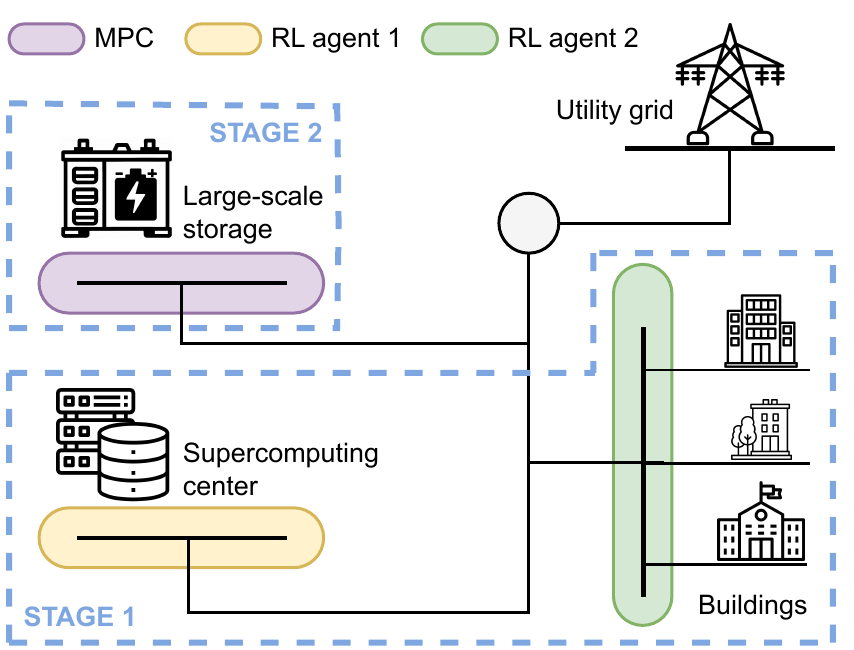}
        }
    \hspace{30pt}
    \subfloat[Centralized control using RL.]{
        \label{fig:opt2}
        \centering
        \includegraphics[width=0.45\textwidth]{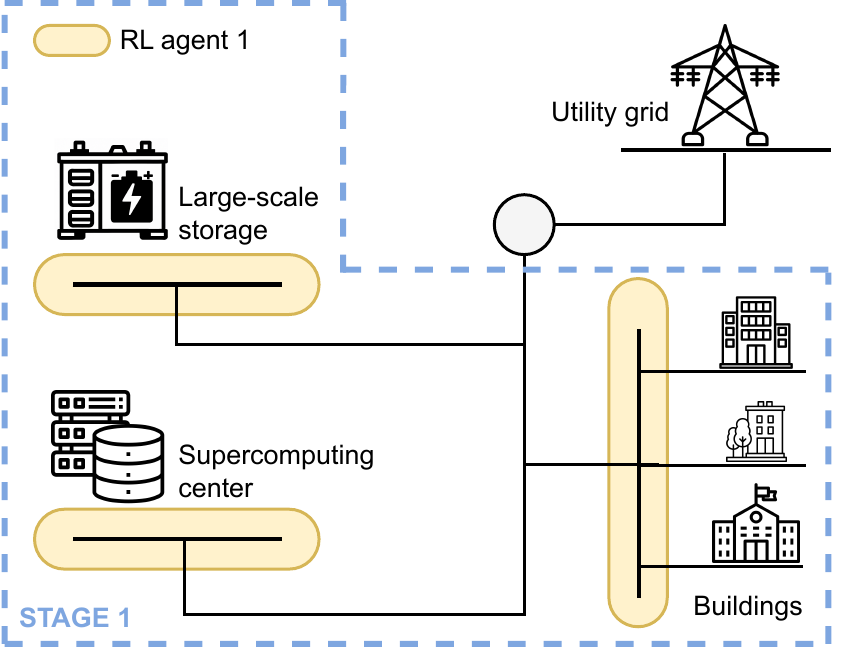}
    }
    \caption{Microgrid of a university campus with several prosumers and different control architectures that can be realized using CommonPower.}
    \label{fig:commonpower_idea}
\end{figure*}

The primary application areas of CommonPower are smart home energy management, demand response, and economic dispatch in modern microgrids, which we model as discrete-time receding-horizon optimal control problems. We consider our system to have a set of buses $\mathcal{N}$,
a state vector $x \in \overline{\mathcal{X}}$, 
an input vector $u \in \overline{\mathcal{U}}$, 
and a disturbance vector $w \in \overline{\mathcal{W}}$, to formulate dynamic and algebraic constraints:
\begin{subequations}\label{eq:sys}
\begin{align}
    \forall i \in \mathcal{N}, \forall t \colon \quad x_{t+1}^i &  = f^i(x_t^i, u_t^i, w_t^i) \label{eq:sys:dynamics} \\
    0 & \leq g^i(x_{[\cdot]}^i, u_{[\cdot]}^i, w_{[\cdot]}^i) \label{eq:sys:ineq_constraints}\\
    0 & \leq d(x_t, u_t, w_t) \label{eq:sys:pf} \\
    w_t & \in \mathcal{W}_t = [\underline{w}_t, \overline{w}_t] \subseteq \overline{\mathcal{W}},
\end{align}
\end{subequations}
where $[\cdot]$ denotes a trajectory and the superscript $i$ indicates association with bus $i \in \mathcal{N}$.
We assume access to forecasts $\hat{w}_t \in \mathcal{W}_t$ of disturbances, where the disturbance set is obtained by conformance checking \cite{roehm2019model}.
The constraints in \eqref{eq:sys:ineq_constraints} typically represent charging or total energy requirements of electric vehicles or flexible loads, which couple multiple time steps and, therefore, consider trajectories.
Power flow constraints between buses are contained in \eqref{eq:sys:pf}.

Generally, the system \eqref{eq:sys} is not exactly known, which is one core motivation for using RL. 
To facilitate optimal control and model-based safeguarding, we assume that a system model $\tilde{f}(\cdot)$, $\tilde{g}(\cdot)$ is available. For optimal control, there are no restrictions on the model complexity, as long as it can be expressed within the symbolic modeling framework Pyomo \cite{hart2011pyomo, bynum2021pyomo} (see Sec. \ref{sec:modeling}).
For providing formal guarantees via our safeguarding approach, we assume that the model $\tilde{f}(\cdot)$, $\tilde{g}(\cdot)$ is obtained via reachset-conformant identification \cite{liu2023guarantees, gruber2023scalable, lutzow2024scalable} to ensure that $x_t \in \mathcal{X}_t$ always holds, where
\begin{equation}\label{eq:reachstep}
\begin{split}
    & \mathcal{X}_{t+1}^i = \tilde{f}^i(\mathcal{X}_t^i, u_t^i, \mathcal{W}_t^i) = \\ &\Big\{ \tilde{f}^i(x_t^i, u_t^i, w_t^i) \, \Big| \,  \exists x_t^i \in \mathcal{X}_t^i, \exists w_t^i \in \mathcal{W}_t^i \Big\}.
\end{split}
\end{equation}
For the sake of computational efficiency, CommonPower currently guarantees this containment condition only if $\tilde{f}^i(\cdot)$ and $\tilde{g}^i(\cdot)$ are input-switched piece-wise continuous functions with piece-wise sign-stable Jacobian. This is further explained in Sec. \ref{subsec:safe_control}.
To guarantee safety beyond the control horizon, we pose the condition that we can always find an input that steers the system into a robust control invariant set  \cite{schaefer2024scalable}. 

\section{Problem Statement}\label{sec:problem_statement}

We consider complex distributed control structures that are becoming increasingly relevant as local generation and energy storage assets enable prosumer-level optimization.
As a motivating example, we introduce an urban university campus microgrid in Fig. \ref{fig:commonpower_idea}. It contains a supercomputing center run by an independent organization, several buildings of the university, and a large-scale battery storage unit.

In our problem statement, the set of all prosumers $\mathcal{P} \subseteq \mathcal{N}$ represents individual stakeholders, while a set of balancing assets $\mathcal{A} = \mathcal{N} \setminus \mathcal{P}$ represents assets that can be utilized by the grid operator to establish grid stability. Prosumers can form disjoint coalitions $\mathcal{G}^k \subseteq \mathcal{P}$, $\bigcap \mathcal{G}^k = \emptyset$, $\bigcup \mathcal{G}^k = \mathcal{P}$ in which members can exchange information, are each controlled by a controller, and are, in a multi-agent setup, mapped to an agent. 

Fig. \ref{fig:opt1} illustrates the high degree of flexibility that can be achieved with this formulation. Here, all buildings and the supercomputing center are prosumers and the storage unit is considered a balancing asset. The three university buildings form the first coalition, and the supercomputing center forms the second coalition, both controlled by one RL agent, respectively.
In this most general setup, the control input for each time step is computed in two stages. In the first stage, all coalitions independently determine their inputs by solving a robust optimal control problem. Since the coalitions do not consider power flow constraints, the grid operator dispatches their balancing assets in the second stage. The imbalance mechanism incurs a cost that can be redistributed to the coalitions in some user-defined way. 
In this work, we refer to \textit{centralized control} for the special case in which there is one coalition $\mathcal{G} = \mathcal{P} = \mathcal{N}$ with a single controller and no balancing assets. This results in a single-stage problem, which is illustrated in Fig. \ref{fig:opt2}.

For the subsequent formalization, let us introduce the nominal trajectory of states $\hat{x}_{[\cdot]}$ under the predicted disturbance trajectory $\hat{w}_{[\cdot]}$.
In every time step and with a control horizon $T$, the two-stage dispatch problem is formalized as follows:
\begin{subequations}
\label{eq:problem}
\begin{align}
    \textbf{Stage 1} \quad \min_{u^{k}_{[\cdot]}} \quad & \sum_{t=0}^{T} J^{k}(u_t^{k}, \mathcal{X}_t^{k}, \mathcal{W}_t^{k}) \\
    \text{s.t.} \quad & \nonumber \\
    \forall i \in \mathcal{G}^k \in \mathcal{G}, \forall t \in [0, T] \quad & \mathcal{X}_{t+1}^i = \tilde{f}^i(\mathcal{X}_t^i, u_t^i, \mathcal{W}_t^i) \label{eq:problem:coalition_constraints} \\
    & 0 \leq \tilde{g}^i(\mathcal{X}_{[\cdot]}^i, u_{[\cdot]}^i, \mathcal{W}_{[\cdot]}^i) \nonumber \\
    \textbf{Stage 2} \quad \min_{u^{\mathcal{A}}_{[\cdot]}} \quad & \sum_{t=0}^{T} J^{\mathcal{A}}(u_t^{\mathcal{A}}, \mathcal{X}_t^{\mathcal{A}}, \mathcal{W}_t^{\mathcal{A}}) \\
    \text{s.t.} \quad & \nonumber \\
    \forall i \in \mathcal{A}, \forall t \in [0, T] \quad & \mathcal{X}_{t+1}^{i} = \tilde{f}^{i}(\mathcal{X}_t^{i}, u_t^{i}, \mathcal{W}_t^{i}) \label{eq:problem:tso_constraints} \\
     & 0 \leq \tilde{g}^{i}(\mathcal{X}_{[\cdot]}^{i}, u_{[\cdot]}^{i}, \mathcal{W}_{[\cdot]}^{i}) \nonumber \\
    & 0 \leq d(\hat{x}_t, u^{\mathcal{A}}_t, \hat{w}_t). \nonumber
\end{align}
\end{subequations}
After each time step, the system evolves according to \eqref{eq:sys} and $\mathcal{X}_0^{(\cdot)} = \{x_0^{(\cdot)}\}$ is a measurement of the current system state.
Note that we only enforce the power flow constraints on the nominal system trajectory for the sake of computational efficiency.
Our formulation can be extended to arbitrarily nonlinear functions and robust power flow feasibility by following the iterative approach from \cite{schurmann2018reachset}. Here, we would compute nominal optimal trajectories with gradually tightening constraints, which are verified under disturbance using reachability analysis based on conservative linearization \cite{althoff2014reachability, althoff2008reachability}.

The problem statement can be considered to be a sequential decision-making process under uncertainty. RL is the standard machine learning approach for solving such problems. 
Decentral control of a system with multiple agents can be realized using multi-agent reinforcement learning (MARL). In MARL, the underlying control problem is commonly modeled as a partially observable Markov game (POMG). It is defined as a tuple $(\mathcal{L}, \mathcal{S}, (\mathcal{U}^\ell, \mathcal{O}^\ell, R^\ell)_{\forall \ell \in \mathcal{L}}, \Phi, \gamma)$ \cite[Sec. 3.1]{yu2022surprising}, where
\begin{itemize}
    \item $\mathcal{L} = \{1,...,L\}$ is the set of agents,
    \item $\mathcal{S} = [\underline{s}, \Bar{s}]$ is the global state space of the environment,
    \item $\mathcal{O}^\ell = [\underline{o}^\ell, \Bar{o}^\ell]$ is the observation space of an agent,
    \item $\mathcal{U}^\ell =[\underline{u}^\ell, \Bar{u}^\ell]$ is the action space of an agent, 
    \item $R^\ell: \mathcal{O}^\ell \times \mathcal{U}^\ell \rightarrow \mathbb{R}$ is the agent-specific reward function, 
    \item $\Phi: \mathcal{S} \times \mathcal{U}_1 \times ... \times \mathcal{U}_L \times \mathcal{S} \rightarrow \mathbb{R}$ is the probability density function modeling state transitions, and
    \item $\gamma \in [0,1)$ is the discount factor used to weigh future rewards.
\end{itemize}
We use the notation $(\mathcal{U}^\ell, \mathcal{O}^\ell, R^\ell)_{\forall \ell \in \mathcal{L}}$ to refer to the tuple of individual quantities $(\mathcal{U}_1,...,\mathcal{U}_L, \mathcal{O}_1,...,\mathcal{O}_L, R_1,...,R_L)$. Single-agent RL control is a special case of the above POMG with $L=1$, resulting in a Markov decision process (MDP). 

The actions of an RL agent $\ell$ controlling one coalition $\mathcal{G}^k$ are $u^\ell_t~=~u^k_t$. The default configuration for observations is $o^\ell_t~=~[ x_t^k, w^k_t, \hat{w}^k_{[t+1, ..., H]}]$, where $H$ is a forecast horizon that can be specified by the user. The default can be overwritten for each agent individually, enabling observation of arbitrary model data of the controlled coalition as well as data from other coalitions. The global state of the POMG is the concatenation of all observations $s_t~=~[o_t^1, ..., o_t^L]$, where duplicates can be removed if desired. Finally, the reward $R_t^\ell~=~-J^k(u_t^k, x_t^k, w_t^k)$ is computed. 

\section{Modeling}\label{sec:modeling}

CommonPower comprises two domains: the object domain and the symbolic domain. The object domain makes it possible to conveniently compose power systems and provides interfaces to external libraries or tools. 
In the symbolic domain, all power system entities have a symbolic representation. They specify the variables and constraints that constitute the system model corresponding to the problem formulation in \eqref{eq:problem}. 
With this approach, object-oriented programming features, such as inheritance, can be leveraged while maintaining full symbolic expressiveness.

\begin{figure}[t] 
\centering
\includegraphics[width=.95\linewidth]{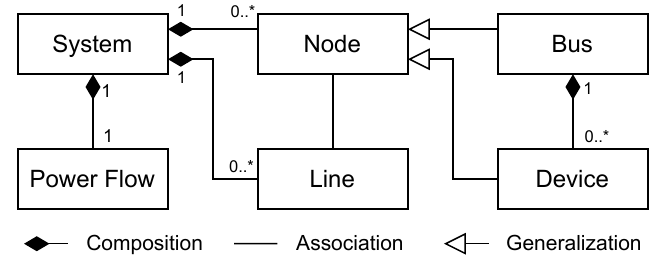}
\caption{UML class diagram of power system entities.}
\label{fig:entities_diagram}
\end{figure}

The class diagram in Fig. \ref{fig:entities_diagram} shows how case studies are composed in the object domain. A number of buses are assigned to the root \texttt{System} object, with each \texttt{Bus} being assigned an arbitrary number of \texttt{Devices}, e.g., batteries, electric vehicles, gas turbines, etc. A singleton instance of \texttt{PowerFlow} representing algebraic constraints between buses is assigned to the system object in conjunction with a number of \texttt{Lines} connecting the buses.
Instances of buses and devices can define the system dynamics $f(\cdot)$ as a simple Python function that is executed during simulation. This realizes the integration of external software, such as Simulink, and supports hardware-in-the-loop simulation.

In the symbolic domain, users can define the overapproximative system model $\tilde{f}(\cdot)$. 
The symbolic modeling is implemented through the framework Pyomo \cite{hart2011pyomo, bynum2021pyomo}. The reasoning behind choosing Pyomo over a more generic symbolic library is that the resulting models can readily be used in optimization problems.  
We would like to highlight that Pyomo can represent arbitrary nonlinear
and mixed-integer constraints, which makes it possible to
model complex dynamics and power flow. The only requirement is that an appropriate solver is chosen.

After having modeled a scenario in the object domain, the system is represented by an object tree of entities, with every entity defining its individual symbolic model. Before running a simulation, CommonPower automatically aggregates this distributed symbolic representation into a global symbolic model in the form of \eqref{eq:problem}. To this end, the object tree is traversed, and all local symbolic models are added to a hierarchical model in which the original tree structure is maintained via Pyomo blocks, as outlined in Algorithm \ref{alg:model_construction}. This hierarchical structure can readily be used for optimization within Pyomo and defines the scope of optimal controllers and safeguards in a decentral control scenario.

\begin{algorithm}[t]
\caption{Global Model Construction}
\label{alg:model_construction}
\begin{algorithmic}[1]
\Procedure{GlobalModel}{$system, horizon, tau$}
    \State Set parameters ($tau$ (sample time), $horizon$)
    \State Create empty global Pyomo model $M$
    \State Create time index set $M.t \gets \{0, 1, \ldots, horizon/tau\}$ 
    
    \For{each node $i$ in system.nodes}
        \State Set node id $nid$ as (parent\_id.class\_prefix + index)
        \State Create Pyomo block $M.nid$
        \State Add node variables $x^i$, $u^i$, $w^i$ to $M.nid$
        \State $\forall t \in M.t\colon$ add constraints for $\tilde{f}^i$, $\tilde{g}^i$\\ \Comment{see \eqref{eq:sys:dynamics}, \eqref{eq:sys:ineq_constraints} (replacing $f^i$, $g^i$ with $\tilde{f}^i$, $\tilde{g}^i$)}

        \For{each child node/device $c$ of $i$}
            \State Recursively add child to model
        \EndFor
    \EndFor
    
    \For{each line $l$ in system.lines}
        \State Set line id $lid$ as (class\_prefix + random suffix)
        \State Create Pyomo block $M.lid$
    \EndFor
    
    \State Add power flow constraints to $M$ \Comment{\eqref{eq:sys:pf}}

    \EndProcedure
\end{algorithmic}
\end{algorithm}

For illustration, the model of a simple battery storage system is presented in Tab. \ref{table:ess_model}. Not all listed model elements need to be manually defined as CommonPower automatically generates them (see Sec. \ref{subsec:modeling_utils}). When instantiating an entity object, the user must configure input and state limits. Furthermore, parameter values need to be passed to the instance.

\begin{table*}[t]
    \centering
    \caption{Battery model. $M$ is a large positive constant.}
    \label{table:ess_model}
    \renewcommand{\arraystretch}{1.5}
    \begin{tabular}{lccccc}
    \toprule
         description & name & type & definition & configuration & constraint expression/domain \\
         \midrule
         active power & $p$ & input & manual & limits & $p \in \mathbb{R}$ \\ 
         charging indicator & $ec$ & state & manual & & $ec \in \{0,1\}$ \\
         cost & $cost$ & state & manual & & $cost \in \mathbb{R}$ \\
         state of charge & $soc$ & state & manual & limits & $soc \in \mathbb{R}_0^+$ \\
         initial soc & $soc^{init}$ & parameter & automatic & value & $soc^{init} \in \mathbb{R}_0^+$ \\
         cost of wear & $\rho$ & parameter & manual & value & $\rho \in \mathbb{R}_0^+$ \\
         charge efficiency & $\eta^c$ & parameter & manual & value & $\eta^c \in [0,1]$ \\
         discharge efficiency & $\eta^d$ & parameter & manual & value & $\eta^d \in [0,1]$ \\
         self-discharge & $\eta^s$ & parameter & manual & value & $\eta^s \in [0,1]$ \\
         indicator constraint 1 & & constraint & automatic & & $\forall t: p_t \geq - M (1-ec_t)$ \\
         indicator constraint 2 & & constraint & automatic & & $\forall t: p_t < M ec_t$ \\
         state initialization & & constraint & automatic & & $soc_0 = soc^{init}$ \\
         dynamics function & & constraint & manual & & $\forall t \in [0,T-1]: soc_{t+1} = \eta^s soc_t + \eta^c (ec_t) p_t + \frac{1}{\eta^d} (1-ec_t) p_t$ \\
         cost function & & constraint & manual & & $\forall t: cost_t = \rho (ec_t) p_t - \rho (1-ec_t) p_t$ \\
    \bottomrule
    \end{tabular}
\end{table*}

\subsection{Modeling Utilities}\label{subsec:modeling_utils}

\subsubsection{Parameter Initialization}

CommonPower automatically creates parameters determining the initial values of all modeled state variables. 
To simulate different parameter values or initial states, users can configure an instance of \texttt{ParamInitializer} instead of some fixed value. Built-in initializers randomly sample values from a given range or loop through a given list of values on every environment reset.

\subsubsection{Piece-wise Expressions}

Piece-wise expressions are often useful in power system modeling, e.g., for the approximation of battery dynamics with piece-wise linear functions \cite{pandvzic2018accurate}. CommonPower models such expressions via mixed-integer constraints and provides a utility that largely automates the generation of corresponding auxiliary constraints based on the big-M method \cite{brown2007formulating}. 
For example, the battery model in Tab. \ref{table:ess_model} represents the dynamic function 
\begin{equation*}
    soc_{t+1} = \eta^s soc_t + \begin{cases}
        \eta^c p_t \quad \text{if} \quad p_t \geq 0 \\
        \frac{1}{\eta^d} p_t \quad \text{otherwise}.
    \end{cases}
\end{equation*}
This case distinction can be modeled via a binary indicator variable $ec$ that takes the value one if $p_t \geq 0$ and zero otherwise. In CommonPower, this variable and corresponding constraints can be straightforwardly defined by invoking the \texttt{MIPExpressionBuilder} utility. The resulting constraints are listed as indicator constraints in Tab. \ref{table:ess_model}. The \texttt{MIPExpressionBuilder} supports the logical operations $\geq$, $>$, $and$, $or$, and $not$.

\subsubsection{Uncertainties}

Users define the symbolic model for the nominal case. Any defined parameter can then be declared uncertain on entity instantiation. Inputs from data providers are assumed to be uncertain as long as the forecaster does not explicitly declare perfect foresight. In case forecasters do not implement a method for obtaining their uncertainty set, CommonPower assumes the smallest possible hyperbox enclosing the forecast and the true value. 

For example, consider the use case of simulating a battery with CommonPower using the model from Tab. \ref{table:ess_model}. Since, in reality, the charging efficiency depends on the state of charge and the temperature \cite{su2023experimental}, one would conduct a system identification based on experimental data. Assume this has shown that the value of the charging efficiency $\eta^c$ lies in the interval $[0.90, 0.99]$ for all relevant conditions. This parametric uncertainty can be injected into the model from Tab. \ref{table:ess_model} by simply declaring $\eta^c$ as uncertain in the battery instance configuration.
CommonPower automatically detects that the uncertainty affects the dynamics function and renders the state of charge uncertain. A more detailed account of how this is considered when solving the system is given in Sec. \ref{subsec:safe_control}.

\subsection{Built-in Entities}

CommonPower includes a range of built-in entities, such as an inflexible load, (curtailable) renewable generators, conventional generators (with rate constraints), energy storage systems, electric vehicles, and a heat pump. 
Built-in buses mainly differ in their modeled cost function, such as maximizing self-sufficiency, minimizing energy cost, or minimizing energy cost in conjunction with carbon intensity. CommonPower provides linearized models that avoid integer variables for many core entities. 
Further, built-in buses can be used to represent external grids or to aggregate nodes in an energy community that minimizes energy cost jointly for all members.
The built-in power flow models represent active power balance, DC power flow constraints \cite[Ch. 6.2.4]{schavemaker_2017_electrical}, and linearized DistFlow \cite{baran1989optimal}. 
Users can use the built-in dynamics model for simulation, implement custom dynamics, inject uncertainties into the existing models, or create entirely custom models.

\subsection{Forecasting}\label{subsec:data_providers}

\begin{figure}[t] 
\centering
\includegraphics[width=0.45\textwidth]{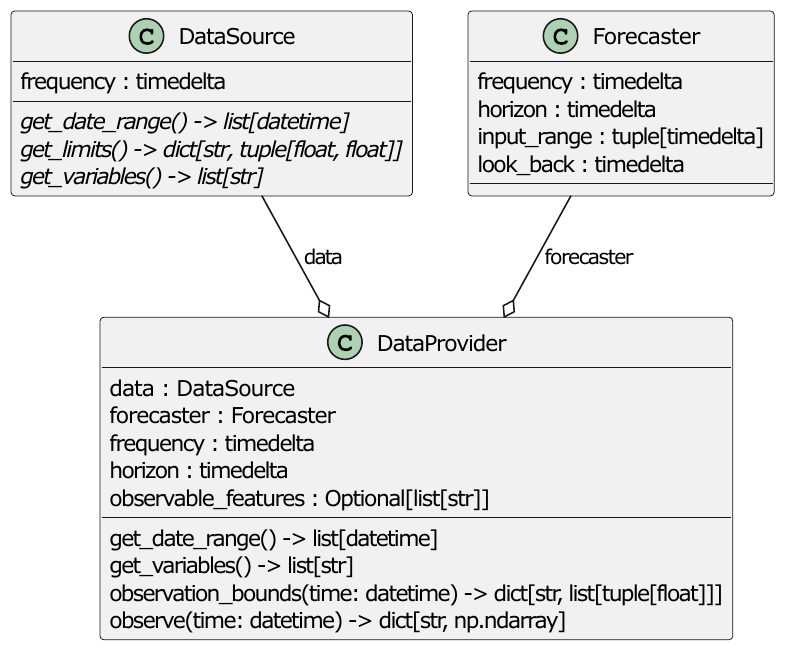}
\caption{UML class diagram of the data provider structure.}
\label{fig:data_provider}
\end{figure}

CommonPower establishes a clear separation between data sources and forecasters and uses a flexible data provider interface. Each disturbance in the system model is required to be associated with an instance of \texttt{DataProvider} that, at each new time step, queries a \texttt{DataSource} for the current value and a \texttt{Forecaster} for predicted future values. A simplified UML class diagram of this structure is provided in Fig. \ref{fig:data_provider}.
Built-in data sources are based on Pandas DataFrames \cite{mckinney-proc-scipy-2010}, CSV files, or cyclically repeating lists of values.
Forecasters return a prediction for the value of a variable for every time step within the control horizon while having access to past and present values of a set of features. 
Some baseline algorithms are built-in, such as perfect forecasts, forecasts with smoothed random noise, or persistence forecasts based on values at certain times in the past.

To facilitate the study of data-driven forecasting approaches, CommonPower implements a framework for training, evaluating, and tuning machine learning models utilizing the Ray Tune library \cite{liaw2018tune}. To obtain a tuned forecasting model on a given data source, the user only needs to select one of the built-in models or provide a custom model implementing PyTorch's \texttt{nn.Module} interface \cite{paszke2019pytorch} with some parameters for the tuning pipeline. The built-in models are configurable standard implementations of a multilayer perceptron (MLP), a long short-term memory network (LSTM), and a transformer.
The pipeline is highly modular and exposes, among others, interfaces for splitting train/test/evaluation sets, extracting data points from the time-series data, and feature/target transformations. The local saving and loading of trained models, including fitted transformations, is managed by CommonPower. 

\section{Control}\label{sec:control}

This section details how control and safeguarding are implemented in CommonPower. We outline the modular architecture and simulation flow, describe the unified interface for heterogeneous controllers, and present our framework for robust safeguarding.

\subsection{High-Level Architecture and Simulation Flow}\label{subsec:architecture}

\begin{figure}[t]
    \centering
    \includegraphics[width=1.\linewidth]{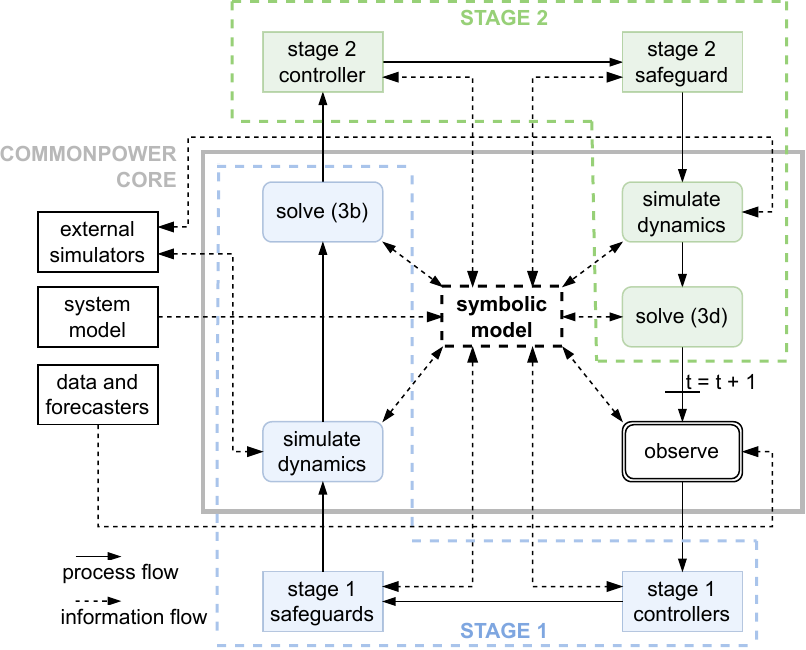}
    \caption{Simplified visualization of one simulation step, highlighting the modularity in CommonPower.}
    \label{fig:commonpower_architecture}
\end{figure}

The architecture of CommonPower is centered around a core module that interacts with the symbolic model and interfaces to all other modules, as visualized in Fig. \ref{fig:commonpower_architecture}. Users can define completely customized controllers and safeguards that can make use of the symbolic model of their controlled entities. Furthermore, adapters to external simulation tools can be integrated into custom or pre-defined components to execute external models. The flexible interface for forecasters and data sources is described in Sec. \ref{subsec:data_providers}.
Overall, our architecture ensures that the symbolic model is consistently managed by the core module while providing maximum flexibility with regard to controllers, safeguards, system simulation, and forecasters.

After the global symbolic model has been synthesized as described in Sec. \ref{sec:modeling}, the system state is initialized as described in Sec. \ref{subsec:modeling_utils}. Each simulation step starts with data sources and forecasters providing values for disturbances, which are updated in the symbolic model.
Next, the controllers of stage~1 (see Sec. \ref{sec:problem_statement}) are queried for their inputs, which are, if necessary, adjusted by safeguards.
All unmodeled dynamics are executed in the ensuing step, possibly integrating external simulation tools. Afterwards, CommonPower solves \eqref{eq:problem:coalition_constraints} for all free variables, i.e., algebraic variables, and states with dynamics that adhere to the symbolic model. This concludes stage 1. In stage 2, balancing assets are dispatched, following the same process as stage 1.

\subsection{Unified Control Approach}\label{subsec:controller_overview}
To enable both the deployment and the training of single-agent and multi-agent systems with heterogeneous controllers, CommonPower offers one unified interface based on the gymnasium API \cite{towers2023gymnasium}, commonly referred to as an \textit{environment}. Developing our own interface was necessary because no standard environment representation has emerged for multi-agent RL thus far. We, therefore, have an internal environment representation and use \texttt{Wrapper} classes to adapt to representations used by the respective RL libraries, such as StableBaselines \cite{raffin2021stable} or MAPPO \cite{yu2022surprising}. 
Our interface realizes the interaction of any control algorithm with the underlying power system. If RL-based controllers are employed, the user can either directly deploy pre-trained policies or train the policies using CommonPower. To use other RL libraries than the above-specified, users can implement a custom \texttt{Runner} class that handles configuration and instantiation of the training process.

\begin{figure}[t]
    \centering
    \includegraphics[width=1.0\linewidth]{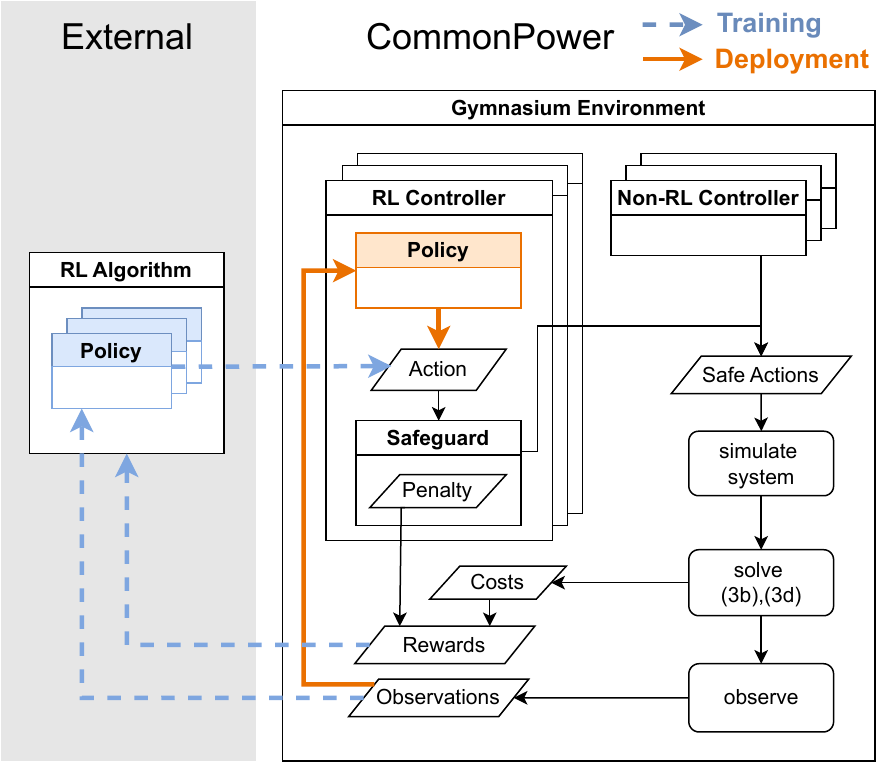}
    \caption{Control flow during training and deployment in CommonPower.}
    \label{fig:commonpower_workflow}
\end{figure}

Fig. \ref{fig:commonpower_workflow} visualizes how our interface handles heterogeneous control structures during training and deployment. It first separates the controllers into RL-based and non-RL controllers. Before training starts, the external RL algorithm has to instantiate one policy for each RL controller. Our interface utilizes the system model to automatically extract the observation and action space for each controller. Please note that while CommonPower supports both discrete and continuous actions, the gymnasium API currently does not support hybrid action spaces. Once training starts, the RL algorithm samples actions from each policy. These can be corrected by the safeguard of the respective controller if required, as described in Sec.~\ref{subsec:safety_shield}. Our interface then collects the actions from all non-RL controllers. The rewards for the RL algorithm are computed for each agent based on the cost function of the controlled entities and the penalty incurred by the safeguarding, as described in Sec.~\ref{subsec:safety_shield}. Finally, observations are obtained by combining states and forecasts.
Storing and loading the trained policies is handled automatically by CommonPower.
During deployment, the same interface can be used to simulate the system with the RL-based and the non-RL controllers.

\subsection{Robustly Safe Control}\label{subsec:safe_control}

We follow an approach that establishes safety by directly including additional constraints in the control problem \eqref{eq:problem}. While this is a computationally efficient strategy, it restricts $\tilde{f}(\cdot)$ and $\tilde{g}(\cdot)$ to input-switched
piece-wise continuous functions with piece-wise sign-stable
Jacobian, as mentioned in Sec. \ref{sec:preliminaries}.
This represents a special case of mixed-monotone functions, for which the reachable set from an uncertainty hypercube can be overapproximated by evaluating the function at two specific vertices of the uncertainty hypercube \cite[Prop. 1]{coogan2015efficient}, establishing a lower and an upper bound for the state trajectory.
Note that this restriction is always fulfilled by any piece-wise linear function and by a large range of device models from the literature. For general non-linear models, approaches such as conservative linearization \cite{althoff2008reachability, althoff2014reachability} can be used to obtain overapproximative linear models.

The uncertainty in the cost function is handled based on scenarios. To this end, we consider every vertex of the uncertainty hypercube of disturbances that are only present in the cost function in addition to the bounds of the state trajectory.  
Instances of the built-in class \texttt{RobustCost} include either the cost of the nominal scenario only, the worst-case scenario only, or a weighted average of all scenarios.

\subsection{RL Safeguarding Framework}\label{subsec:safety_shield}
Through its modular architecture, CommonPower enables the integration of various safeguarding approaches, for example, using Lyapunov stability theory \cite{cui2022decentralized} or barrier functions \cite{vu2021safe}. However, leveraging our symbolic system representation makes it possible to automate the synthesis of safeguards that are a generalized form of the predictive safety filter from \cite{wabersich2021predictive}. Specifically, the safeguards find a safe action $u^\ell_t$ based on the proposed action $a^\ell_t$ of RL agent $\ell$ by solving
\begin{equation}
\label{eq_projection}
\begin{split}    
    & u^\ell_t = \argmin_{u^\ell_t} \phi(u^\ell_t, a^\ell_t) \\
    \text{s.t.} \quad & \eqref{eq:problem:coalition_constraints},
\end{split}
\end{equation}
where $\phi(u^\ell_t, a^\ell_t)$ is a cost function. Built-in classes of \texttt{Safeguard} implement two common strategies. With $\phi(u^\ell_t, a^\ell_t) = \|u^\ell_t - a^\ell_t\|_2$, the approach is commonly called action projection \cite[Eq. (12)]{gros2020safe}. In contrast, with $\phi(u^\ell_t, a^\ell_t) = 0$, we obtain random actions with the solver initialization as the source of randomness. This is an example of so-called action replacement \cite[Sec. 2.1]{krasowski2022provably}. We refer the reader to the study in \cite{krasowski2022provably} for additional theoretical background and an experimental evaluation of different model-based safeguarding approaches.

We make the standard assumption of well-posedness of our problem, implying the existence of safe actions. To relax this assumption, one could implement a custom safeguard with a fallback action such as load shedding. 
However, the question of whether we can always find safe actions in polynomial time is challenging, since there exists no method to do so for general nonlinear problems. For QPs with a convex objective function, as in the case of projection, we can only guarantee this if the constraints define a convex feasible set. This can be achieved by using linearized system dynamics as well as linear power flow models or convex inner approximations of the power flow feasible set \cite{lee2019convex, lee2021robust}.

When using safety filtering during RL training, the behavior policy used to gather data differs from the target policy that is being learned. Normally, a policy is updated based on a batch of tuples $(o^\ell_t, a^\ell_t, o^\ell_{t+1}, r^\ell_t)$. When a correction of $a^\ell_t$ becomes necessary, this tuple could be changed to $(o^\ell_t, u_t^\ell, o^\ell_{t+1}, r^\ell_t)$ which features the safe action and the reward obtained from applying this action. However, this would mean updating the policy with actions that do not stem from the most recent policy, which is an expected procedure for off-policy algorithms but can be problematic for on-policy algorithms \cite{krasowski2022provably}. Instead, we add an adaption penalty to the reward 
\begin{equation}
    \tilde{r}^\ell_t = R^\ell(o^\ell_t, u^\ell_t, o^\ell_{t+1})+R^{\ell, pen}(a^\ell_t, u_t^\ell)
    \label{eq:safe_reward}
\end{equation} such that the tuple used for learning is $(o^\ell_t, a^\ell_t, o^\ell_{t+1}, \tilde{r}^\ell_t)$. This penalty informs the agent about the intervention of the safeguard. CommonPower includes two built-in implementations of this \texttt{SafetyPenalty}; one represents the Euclidean distance between action and safe input, and the other is a configurable constant penalty if the action was adjusted.  

\section{Experiments}\label{sec:experiments}
We present four experiments that demonstrate the capabilities and illustrate potential research directions that are enabled by the unique features of CommonPower and could not have been readily realized with any existing tool. All experiments were conducted on an Intel Core i9-14900K.

\subsection{Importance of Robust Safeguarding}\label{sec:exp_robustness}
For the first two experiments, we study a building management task, where the goal is to control the power set points of a heat pump and a battery storage system so that the electricity cost of the building is minimized while the temperature is kept close to a chosen set point. Designing a safe controller for this task is challenging because of the inertia of the indoor temperature.

The building has an inflexible active power consumption that always has to be satisfied, and a photovoltaic generator.
We model the battery dynamics as in Tab. \ref{table:ess_model} with $\eta^s = 0$ and $\eta^c=\eta^d=1.0$. The dynamic equations for the heat pump are taken from \cite{bianchi2006adaptive}. As the ground truth data for the PV generation and active power consumption, we use the Simbench \cite{meinecke2020simbench} dataset \textit{1-LV-rural2--1-sw}\footnote{\url{https://simbench.de/de/datensaetze}}. For the outdoor temperature and the coefficient of performance of the heat pump, we use the When2Heat dataset \cite{ruhnau2019time}.

\begin{figure}[t]
    \centering
    \includegraphics[width=1.\linewidth]{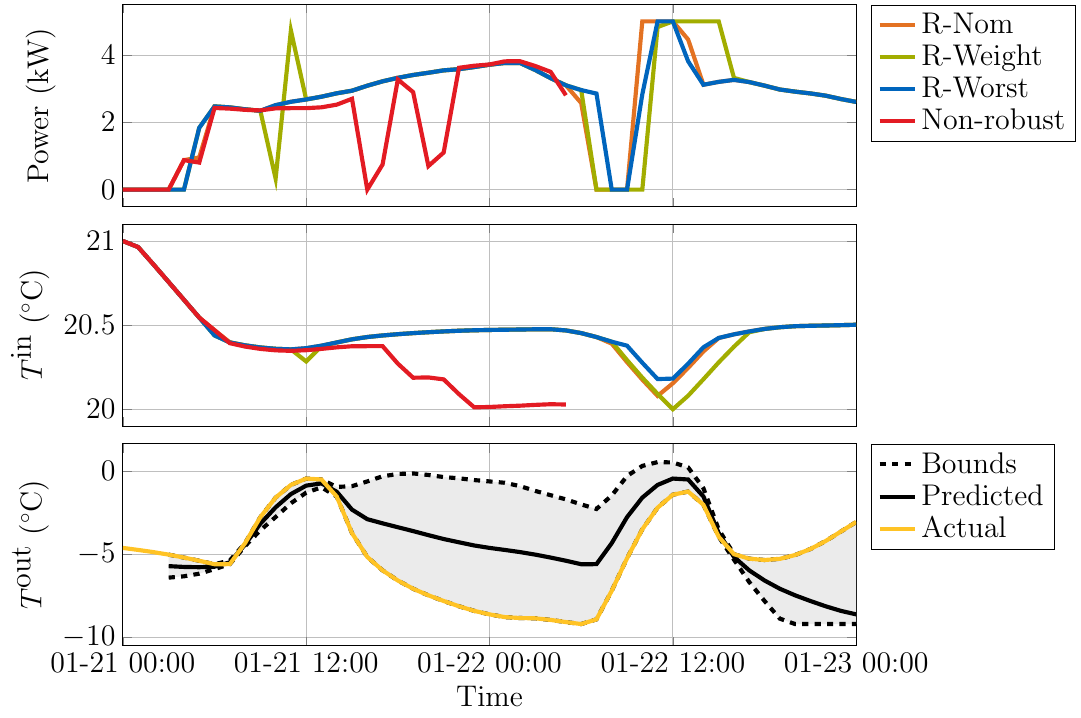}
    \caption{Comparison of robust and non-robust model predictive control for a household featuring a heat pump and a battery storage system.}
    \label{fig:robust_heatpump}
\end{figure}
We first show the necessity of utilizing the robust safeguarding described in Sec. \ref{subsec:safe_control}. To this end, we simulate the above-described system for $48$ hours with a time interval of one hour. We limit the indoor temperature to the interval $T^\textnormal{in}~\in~[20\degree C, 22\degree C]$ with the target indoor temperature set to $21\degree C$. The outdoor temperature $T^\text{out}$ is predicted using the values from the previous day.

Fig. \ref{fig:robust_heatpump} compares a naive optimal control approach to robust optimal control with the three different built-in cost functions: the nominal cost for the realized scenario, a weighted sum of all scenarios, and the cost for the worst-case scenario. Since the outdoor temperature is overestimated for a long period of time, the naive optimal controller is not able to keep the indoor temperature within the desired range, and the simulation fails after roughly 30 time steps. In contrast, all robust controllers can keep the system within the limits. The controller optimizing the worst-case cost results in the smallest deviation from the desired temperature set point. The overall cost is the highest when using the weighted sum of all scenarios ($58.86$) and lowest using the nominal cost ($58.22$).

\subsection{Comparison of Single- and Multi-agent RL}\label{sec:exp_sarl_marl}
\begin{table*}[t]
    \centering
    \caption{Comparison of single-agent RL (PPO) and multi-agent RL (IPPO, MAPPO).}
    \label{tab:comparison_single_multi}
    \begin{tabular}{@{}llccccccccc@{}}
    \toprule
    \multicolumn{1}{c}{} & \multicolumn{1}{c}{\textbf{}} & \multicolumn{3}{c}{\textbf{PPO}}      & \multicolumn{3}{c}{\textbf{IPPO}} & \multicolumn{3}{c}{\textbf{MAPPO}} \\ 
    \cmidrule(lr){3-5} \cmidrule(lr){6-8} \cmidrule(l){9-11}
    \multicolumn{1}{c}{} & \multicolumn{1}{c}{\textbf{}} & Safe     & Safe-Pen & Unsafe & Safe & Safe-Pen & Unsafe & Safe  & Safe-Pen & Unsafe \\ 
    \cmidrule(lr){3-5} \cmidrule(lr){6-8} \cmidrule(l){9-11}
    \multicolumn{1}{c}{} &                               & \multicolumn{9}{c}{Training}                                               \\ \midrule
    \textbf{Interventions}        & Mean                          & $15,238$  & $2,751$   & n.a.   &  $23,140$  & $1,665$   & n.a.   &  $24,862$ & $1,664$   & n.a.   \\
    \textbf{Safeguard time}       & Mean                          & $14.3\%$ & $14.5\%$ & n.a.   & $12.8\%$ & $13.3\%$ & n.a.   &  $13.0\%$   & $13.3\%$ & n.a.   \\ \midrule
    \multicolumn{1}{c}{} &                               & \multicolumn{9}{c}{Deployment}                                             \\ \midrule
    \textbf{Costs}                & Min                           & $9.21$   & $8.59$   & $9.36$ & $8.77$  & $8.20$   & $8.99$     & $8.81$  & $8.24$   &  $9.03$   \\
    \textbf{}            & Max                           & $10.05$   & $9.82$   & $9.56$ & $9.57$  & $9.66$   & $8.99$     & $9.22$  & $9.01$   &  $11.12$  \\
    \textbf{}            & Mean                          & $9.61$   & $9.04$   & $9.46$ & $9.10$ & $8.94$   & $8.99$     &  $9.01$  & $8.71$   &  $10.08$  \\
    \textbf{Failed seeds}         & -                             & 0        & 0        & 3      &   0  &    0     &    4    &    0   &      0    &    3    \\ \bottomrule
    \end{tabular}
    \end{table*}
The task described in Sec. \ref{sec:exp_robustness} can be solved using a centralized control structure -- one controller for both heat pump and battery -- or a decentralized one. Centralized control, realized with single-agent RL, has the advantage of having access to the full observation space. On the other hand, with multi-agent RL, the task of each individual agent is less complex. CommonPower enables an effortless comparison of both approaches for a given task.

\begin{figure}[t]
    \centering
    \includegraphics[width=\linewidth]{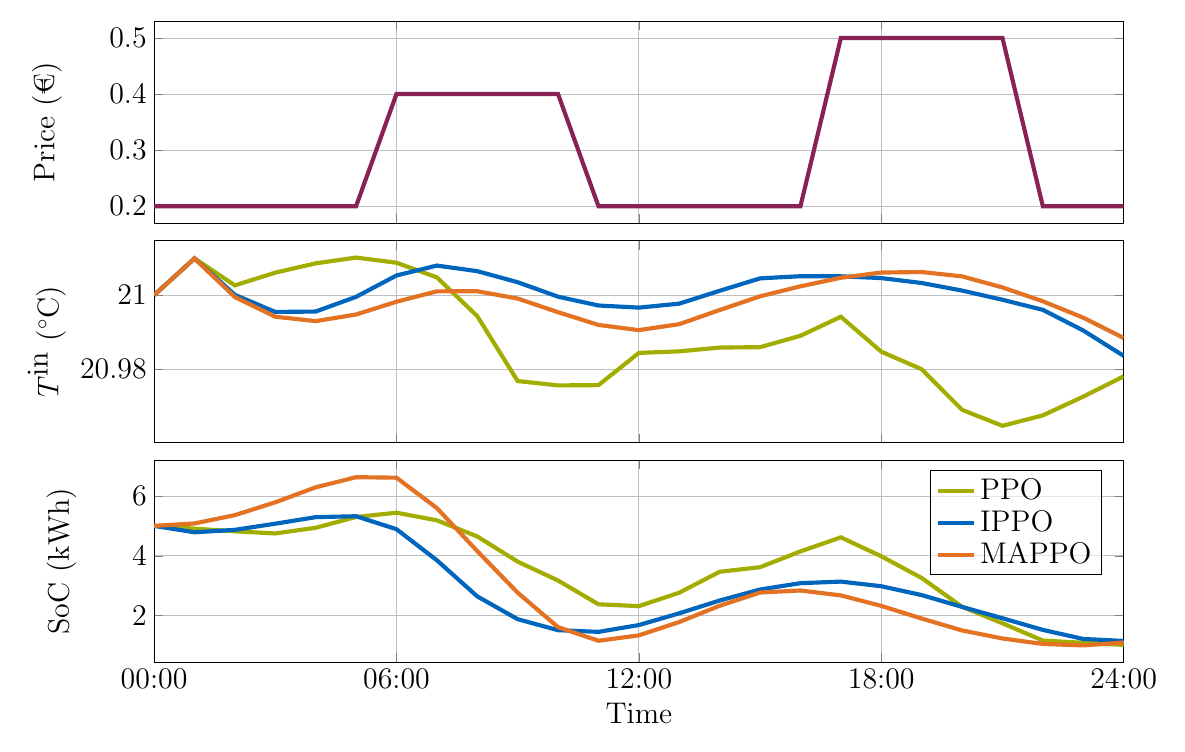}
    \caption{Centralized (PPO) and decentralized control (IPPO, MAPPO) under time-of-use electricity prices for a household featuring a heat pump and a battery storage system.}
    \label{fig:comparison_single_multi}
\end{figure}

The system setup is very similar to the one described in Sec. \ref{sec:exp_robustness}.
We relax the constraints for the indoor temperature such that $T^{\textnormal{in}} \in \left[18\degree C, 24\degree C\right]$. To reduce noise, we employ perfect forecasts for all quantities in this experiment. Finally, we use time-of-use electricity prices with the profile shown in Fig. \ref{fig:comparison_single_multi}. 

\begin{figure*}[ht]
    \centering
    \subfloat[LSTM architecture]{
        \label{fig:tuning_LSTM}
        \centering
        \includegraphics[width=0.47\textwidth]{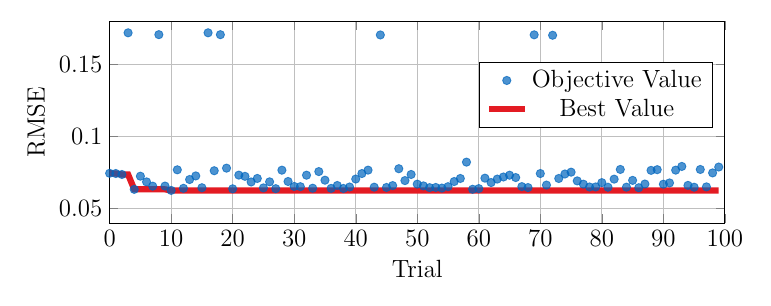}
        }
    \quad
    \subfloat[Transformer architecture]{
        \label{fig:tuning_transformer}
        \centering
        \includegraphics[width=0.47\textwidth]{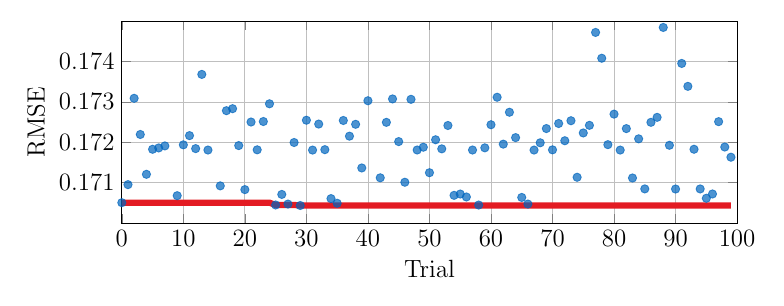}
    }
    \caption{Root mean squared error (RMSE) over the test data set of active power consumption for 100 hyperparameter trials.}
    \label{fig:tuning}
\end{figure*}

During single-agent RL training, we minimize the deviation of the indoor temperature from the set point $T^\text{set}$, the battery degradation, and the electricity cost, realized by
\begin{equation*}
    r_t = - (p_t^\textnormal{grid} \varphi_t + c^\textnormal{comfort} (T^\textnormal{in} - T^\text{set})^2 + c^\textnormal{degrad}  |p_t^\textnormal{battery}|).
\end{equation*}
Here, $p^\textnormal{grid}$ is the power the household has to draw from the external grid, $p^\text{battery}$ is the charge/discharge power of the battery, $\varphi$ is the electricity price, and $c^\textnormal{comfort}, c^\textnormal{degrad}$ are weighting factors for the comfort cost and the battery degradation cost. The observation of the agent is $o_t~=~\left[\textnormal{soc}, p^\textnormal{PV}_{[\cdot]}, p^\textnormal{load}_{[\cdot]}, T^\textnormal{in}, z^\textnormal{hp}, T^\textnormal{out}_{[\cdot]}, \varphi_{[\cdot]} \right]$, where $z^\textnormal{hp}$ represents the internal states of the heat pump and $p^\textnormal{PV}, p^\textnormal{load}$ are the power of the PV array and the non-controllable load, respectively.

For multi-agent RL, we split the electricity cost between the agent controlling the battery and the one controlling the heat pump, realized by
\begin{align*}
    r_t^\textnormal{battery} &= - \left(\frac{1}{2}p_t^\textnormal{grid} \varphi_t + c^\textnormal{degrad}  |p_t^\textnormal{battery}| \right), \\
    r_t^\textnormal{heatpump} &= - \left(\frac{1}{2}p_t^\textnormal{grid} \varphi_t + c^\textnormal{comfort} (T^\textnormal{in} - T^\text{set})^2 \right).
\end{align*}
This means that the reward for the individual agent is non-stationary as it also depends on the actions of the other agent. 
The agents receive
\begin{align*}
    o_t^\textnormal{battery} &= \left[\textnormal{soc}, p^\textnormal{PV}_{[\cdot]}, p^\textnormal{load}_{[\cdot]}, T^\textnormal{in}, T^\textnormal{out}_{[\cdot]},\varphi_{[\cdot]} \right], \\
    o_t^\textnormal{heatpump} &= \left[T^\textnormal{in}, z^\textnormal{hp}, T^\textnormal{out}_{[\cdot]}, \varphi_{[\cdot]} \right]
\end{align*}
as observations, respectively.

We compare the performance of PPO \cite{schulman2017proximal} as a single-agent RL algorithm with the two multi-agent algorithms IPPO \cite{dewitt2020independent} and MAPPO \cite{yu2022surprising} on a given day. The difference between the two latter algorithms lies in the observation space of the critic used to guide the policy during training: In MAPPO, it is conditioned on the union of the individual observations as a means for alleviating non-stationarity. We compare safe training using action projection and a proportional penalty $r^{\ell, pen}_t = 10 \, \|a^\ell_t - u^\ell_t\|$ to safe training without a penalty to analyze how the penalty affects performance across the different algorithms. Furthermore, we report the results for training without a safeguard (unsafe). Here, we cannot employ proportional penalties as safe actions are not computed. Therefore, to inform the agent about constraint violations, we use a constant penalty $r^{\ell, pen}_t = -50$. Training is performed over data from one day with constant initial states of the battery and the heat pump to limit training times. During deployment, we perturb the initial state of the battery to achieve a different data distribution compared to training. For all algorithms, we first perform hyperparameter tuning over 100 trials for the batch size, the learning rate, and the initial standard deviation, and then use the best hyperparameters to train on five different random seeds. 

Tab. \ref{tab:comparison_single_multi} shows that the agents trained with a projection safeguard and a proportional penalty consistently deliver the best average performance during deployment. MAPPO delivers the best overall result, which can be attributed to the reduced complexity in the action space compared to PPO and the measures against non-stationarity explained above. Fig. \ref{fig:comparison_single_multi} shows the best-performing seed for each approach. We see that all learn to exploit low-level prices to charge the battery and pre-heat the building. 
Please note that unsafe training leads to failures during deployment for all algorithms, as shown in Tab. \ref{tab:comparison_single_multi}, highlighting the necessity of using a safeguard. This also justifies the computational overhead incurred by the safeguarding, which constitutes $13.5\%$ of the training time on average.  

\subsection{Effects of Forecaster Choice on Dispatch Cost}\label{sec:exp_3}

The topology of our test system is based on the \textit{Kerber Landnetz Kabel 2} network taken from the pandapower library \cite{thurner2018pandapower}.
The network has 30 buses, 14 of which are households, and is connected to the external grid via a substation. The topology is imported from pandapower and all parameters are maintained, e.g., line admittances. We use the DC power flow model. 
During import, devices are added to households in a stochastic fashion. Each household has an inflexible load with a probability of 100\%, a battery with a probability of 50\%, and a photovoltaic generator with a probability of 50\%.
The ground truth generation/load profiles are again based on the Simbench dataset \textit{1-LV-rural2--1-sw}.

We simulate the system described above with a centralized optimal controller over one year. The forecast horizon of the optimal controller is set to 12 hours with a frequency of one hour. We assume a constant buying ($0.37$ €/kWh) and selling ($0.08$ €/kWh) price.

\begin{figure*}[ht]
    \centering
    \subfloat[LSTM architecture]{
        \label{fig:error_LSTM}
        \centering
        \includegraphics[width=0.47\textwidth]{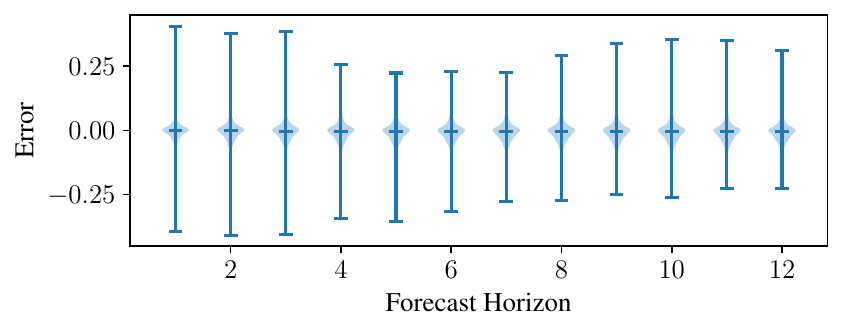}
        }
    \quad
    \subfloat[Transformer architecture]{
        \label{fig:error_transformer}
        \centering
        \includegraphics[width=0.47\textwidth]{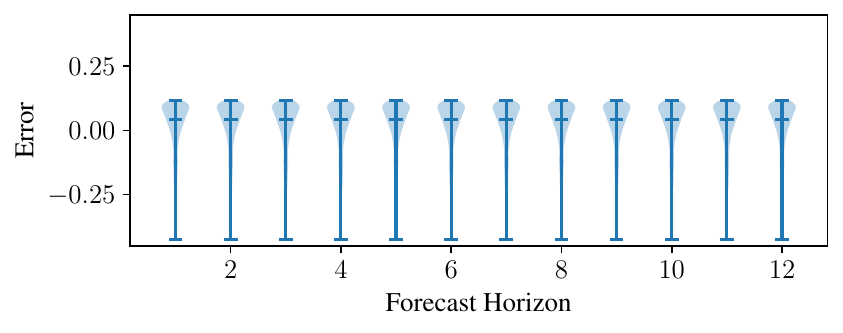}
    }
    \caption{Error distribution during deployment between the true and the forecasted load for each time step in the forecast horizon. The results are averaged over all households.}
    \label{fig:error_dists}
\end{figure*}

Our goal is to analyze the effect of forecast accuracy on the yearly electricity cost of the system. For simplicity, we choose perfect forecasts for the photovoltaic generation. The only uncertainty thus stems from the forecasts for the active power consumed by each household. They are provided by either a pre-trained LSTM model or a pre-trained transformer. We use CommonPower to first tune hyperparameters (batch size, dropout rate, number of layers, learning rate, lookback window) and then train the forecasting model with the identified set of hyperparameters. The tuning is performed over 100 trials with an 80-20 split for training and validation data. Fig.~\ref{fig:tuning} shows the spread of the root mean squared error used as the optimization objective during tuning. While the transformer architecture seems to be more robust, the LSTM delivers substantially better forecasting accuracy during training. However, the controller using the LSTM forecasts obtains a higher yearly electricity cost ($4220.51$€) than the one using the forecasts from the transformer ($4206.23$€). Fig.~\ref{fig:error_dists} visualizes the error distributions of the two forecasters. We observe that the transformer-based forecaster always predicts very similar values over the forecast horizon, leading to overestimating the anticipated load in most cases. As a result, the controller has to be more conservative when allocating resources, which explains the improved closed-loop performance.

\subsection{Influence of Modeling Choices on Computation Time}

This experiment investigates the computation time of a centralized MPC with perfect forecasts across three different network sizes, two power flow models, and linear as well as piecewise-linear ESS models. The setup is identical to Sec. \ref{sec:exp_3} except that every household is assigned an inflexible load, photovoltaic generation, and a battery storage system. We simulate one day and average the computation time over all time steps. The results are listed in Tab. \ref{tab:comparison_scaling}.  

We can observe that under power balance constraints, the computation time scales approximately linearly with the network size. The use of a piecewise-linear ESS model increases the computation time by roughly 50\%. 
Under LinDistFlow constraints, we obtain significantly higher computation times with the piecewise-linear ESS model. On a technical level, this is due to the fact that the presolve routine of the solver cannot eliminate as many variables as under power balance constraints, since they are more closely coupled. For reference, the optimization problem for the largest network has 24567 continuous variables, 1898 binary variables, and 1752 quadratic objective terms.
Please note that this is only an exemplary measurement and cannot be generalized, since the time to solve mixed integer programs depends strongly on the problem structure and several solver parameters.  

\begin{table}[]
\centering
\caption{Simulation time [s] per time step.}
\label{tab:comparison_scaling}
\begin{tabular}{lccc}
\toprule
Buses/Households & 30/14 & 116/57 & 294/146 \\
\midrule
power balance + linear ESS  & 0.13    & 0.46    & 1.33   \\
power balance + piecewise-linear ESS   & 0.17    & 0.75    & 2.00    \\
LinDistFlow + linear ESS  & 0.17    & 0.75    & 2.17    \\
LinDistFlow + piecewise-linear ESS   & 3.63    & 5.58    & 12.38    \\
\bottomrule
\end{tabular}
\end{table}

\section{Conclusion}\label{sec:conclusion}
We introduce CommonPower, a comprehensive toolbox designed as a one-stop-shop solution for the modeling and simulation of safe controllers in smart grids.
Leveraging a symbolic representation of the system, model-based RL safeguards as well as robust model predictive controllers are automatically derived, significantly accelerating case study design for practitioners.
CommonPower features a flexible coalition-based approach, admitting the complex distributed control structures of modern and future smart grids.
The modular software architecture exposes, among others, unified interfaces for single-agent and multi-agent RL algorithms, external simulation tools, data sources, and machine-learning-based forecasting models. 
Due to this high amount of flexibility, CommonPower can easily be integrated into existing projects, providing a common foundation for a vast variety of use cases and paving the way for increased real-world adoption of data-driven smart grid control.
Planned extensions are the implementation of further entity models, hierarchical control, contingency constraints, further safeguarding approaches, and explicit modeling of energy trading.

\bibliography{root}
\bibliographystyle{ieeetr}




\end{document}